\documentclass{PoS}

\usepackage{slashed}
\usepackage{amssymb,amsmath,latexsym}
\usepackage{graphicx,ctable,booktabs,verbatim}
\usepackage{float}
\usepackage{setspace}
\usepackage{multicol}
\usepackage{mathtools}

\newcommand{\defeq}{\vcentcolon=}

\newcommand\be{\begin{eqnarray}}
\newcommand\ee{\end{eqnarray}}

\title{Second-Order Fermions and the Standard Model}

\ShortTitle{Second-Order Fermions and the Standard Model}

\author{\speaker{Johnny Espin}\thanks{Work done in collaboration with Kirill Krasnov.}\\
        School of Mathematical Sciences\\
 			University of Nottingham\\
        E-mail: \email{pmxje3@nottingham.ac.uk}}

\abstract{We present a construction of a non-hermitian fermionic Lagrangian which has a second-order kinetic term. Despite the non-hermicity of the latter, the theory is unitary and the perturbation theory that can be derived is equivalent to the usual one derived from a first-order formalism. Having this in mind, the construction of a second-order Standard Model allows a more compact description of the theory. This work is based on \cite{Espin:2013wia,paperuni} and citations therein.}

\FullConference{Proceedings of the Corfu Summer Institute 2014 "School and Workshops on Elementary Particle Physics and Gravity",\\
		3-21 September 2014\\
		Corfu, Greece }

\begin{document}

\section{Introduction}
Since the emergence of Quantum Mechanics (QM), much effort was put into finding relativistic wave equations that would govern the dynamics of mechanical systems. Schr\"odinger and then Klein and Gordon formulated a second-order wave equation, but at that time it seemed that the nature of the latter violated some fundamental properties of mechanical systems: the Klein-Gordon wave admitted a positive and a negative energy solution.  British physicist Paul A.M. Dirac believed that the issue relied on the second-order nature of the differential equation. He therefore sought a first-order differential equation that respected the relativity principle. His theory was formulated in 1928 and the Dirac equation was later shown to describe relativistic spin $1/2$ particles: fermions. This was followed by the development of Quantum Electrodynamics (QED), the relativistic quantum theory of light and matter interactions which was then generalised into Yang-Mills (YM) theory, the theory of non-abelian $SU(n)$ gauge fields. This summarises the success of particle physics in the last century, success that culminated with the edification of the Standard Model (SM) of particle physics.
Yet something can be seen as puzzling. Indeed, one of Dirac's reasons to seek a first-order differential equation was the misinterpretation of the negative energy solutions. However immediately after the discovery of his equation, it became clear that Nature admitted particles and antiparticles (positive and negative energy solutions). Nonetheless, fermions remained the only dynamical system that only admitted a first-order description. Indeed, both General Relativity and Yang-Mills theory admit a first- and second-order formulation \cite{Deser:1987uk,Martellini:1997mu}. As for fermions, because their equation is derived from the relativity principle, it also satisfies the Klein-Gordon equation; its Lagrangian is, however, first order in derivatives. Hence, it is natural to ask whether a (fully) second-order formulation of fermions that contains all the information of the original one is possible.
\\
Let us make the last argument more precise. A Dirac spinor can be written as the sum of two unitary infinite dimensional representations of the Lorentz group $SO(3,1)$ (or its double cover $SL(2,\mathbb{C})$):

\begin{equation}
\Psi_D \in \left(1/2,0 \right) \oplus \left( 0,1/2 \right)
\end{equation}
which we call left-(unprimed) and right-handed (primed) respectively. The Dirac equation is then derived from the Dirac Lagrangian, here in $3+1$ dimensions with the metric $\eta_{\mu\nu}=(-,+,+,+)$:

\begin{equation}
\mathcal{L}_D= \bar\Psi_D \left(-i\slashed \partial  - m\right)\Psi_D\label{dirac}
\end{equation}
with $\slashed \partial = \gamma^\mu \partial_\mu$ and we have the algebra of (Dirac) gamma matrices:

\begin{align}
\{ \gamma^\mu, \gamma^\nu \}= -2\eta^{\mu\nu}, \quad \left( \gamma^\mu\right)^\dagger=\gamma^0 \gamma^\mu \gamma^0, \quad \gamma_5 = i\gamma^0\gamma^1\gamma^2\gamma^3,\quad \left( \gamma_5\right)^\dagger=\gamma_5
\end{align}

This Lagrangian generalises in a straightforward way so as to include the interaction of fermions and photons. We see that the Dirac equation

\begin{equation}
\left(-i\slashed \partial  - m\right)\Psi_D=0
\end{equation}

relates spinors of one chirality to the other through the off-diagonal entries of the Dirac matrices\footnote{In the case of Majorana fermions the spinor is linked to its hermitian conjugate through the Dirac equation.}. This is heuristically why a second order Lagrangian of the type
\begin{equation}
\mathcal{L}_D ~\stackrel{?}{=} ~ \bar\Psi_D \left(-\square +m^2\right)\Psi_D
\end{equation}
does not work since the Klein-Gordon operator is diagonal and hence we lose information contained in the Dirac equation. 

A lot of insight can be gained on the issue when one expresses all the quantities in terms of two-component spinors (see for example \cite{Dreiner:2008tw}). In this paper, we will construct the simplest second-order spinor field theories; those of a (Weyl-)Majorana and of a Dirac fermion. Then, we will apply the same construction to the Standard Model. Finally, we will shortly discuss Perturbation Theory and Unitarity in this framework.

\section{Second-order (Weyl-)Majorana and Dirac fermions}
We start by considering the much simpler setup of a single (Weyl-)Majorana spinor and then repeat the procedure for a Dirac spinor.

\subsection{First-order Lagrangian}

The Lagrangian for a single Majorana fermion reads:
\be\label{weyl-massless}
{\cal L}_{\text{Maj}}=  -i\sqrt{2} \lambda^\dagger \partial \lambda- (m/2)\lambda\lambda - (m/2)\lambda^\dagger\lambda^\dagger
\ee
Here $\lambda_A$ is a two-component spinor, $\lambda_A^{\dagger}$ is its Hermitian conjugate  and $\theta_\mu^{A'A}$ is the soldering form
\be
\theta^A_{\mu A'} \theta_{\nu A}{}^{A'} = \eta_{\mu\nu}
\ee
where $\eta_{\mu\nu}={\rm diag}(-1,1,1,1)$. We have also introduced the notation $\partial^{AA'} := \theta^{\mu\,AA'} \partial_\mu$ and written the Lagrangian in an index-free way. Our index-free convention is that the unprimed fermions are always contracted as $\lambda\chi\equiv \lambda^A \chi_A$, while the primed fermions are contracted in the opposite way $\lambda^\dagger \chi^\dagger\equiv (\lambda^\dagger)_{A'}(\chi^\dagger)^{A'}$.

\subsection{Equations of motion and second-order theory}
The (first-order) equations of motion for the unprimed spinor are:

\begin{align}
-i\sqrt{2}\partial \lambda -m \lambda^\dagger =0
\end{align}

The second-order formulation can then be obtained by carrying out the Berezin path integral over the primed spinors (which are treated as independent variables), which effectively amounts to substituting the above equation of motion in the Lagrangian. Then we have:

\begin{align}
\label{weyl-massless2}
{\cal L}_{\text{Maj}}=  -\frac{1}{m} \partial \lambda\partial \lambda- (m/2)\lambda\lambda
\end{align}
which is equivalent, after an appropriate rescaling of the fields to:

\begin{align}
{\cal L}_{\text{Maj}}=  -\partial \lambda\partial \lambda- (m^2/2)\lambda\lambda
\end{align}

The second-order Weyl theory is then simply obtained from the massless limit (after the path integral of the primed spinors has been solved):

\begin{align}
{\cal L}_{\text{Weyl}}=  -\partial \lambda\partial \lambda
\end{align}

\subsection{Dirac fermions}

In order to describe Dirac fermions, we consider two massive Majorana fermions of equal mass. The system is then invariant under and internal ${\rm SO}(2)$ symmetry. Since ${\rm SO}(2)\sim {\rm U}(1)$, we can introduce a complex linear combination of the spinors and make the second symmetry explicit. It can then be made local by introducing a ${\rm U}(1)$ gauge field and by minimally coupling the fermions. Thus, we define
\be\label{cov-ders}
D \xi = (\partial - i e A)\xi, \qquad
D\chi = (\partial + i e A)\chi,
\ee
where $A^{AA'}= \theta^{\mu\,AA'}A_\mu$ is the electromagnetic potential and $e$ is the electron charge. The gauge transformation rules are: for the fermions $\xi\to e^{i\alpha} \xi, \chi\to e^{-i\alpha}\chi$, and for the electromagnetic potential $A_\mu\to A_\mu-(1/e) \partial_\mu \alpha$. The Lagrangian becomes
\be\label{dirac-a}
{\cal L}_{\text{Dirac}}=- i \sqrt{2} \xi^\dagger D \xi-i \sqrt{2} \chi^\dagger D \chi - m\chi\xi -m\xi^\dagger \chi^\dagger,
\ee
where as before $D:=\theta^\mu D_\mu$. In order to obtain the second-order Lagrangian, the same procedure is applied to both primed spinors, and we obtain

\be\label{eqs-dirac}
{\cal L}_{\text{Dirac}} = - 2 D \chi  D \xi  - m^2 \chi \xi,\quad \xi^\dagger = -\frac{i\sqrt{2}}{m} D \chi, \qquad \chi^\dagger = -\frac{i\sqrt{2}}{m} D \xi.
\ee

\section{Short review of the Standard-Model}
For simplicity, we only consider the quark sector of the SM, for the complete picture, we refer the reader to \cite{Espin:2013wia}.

\subsection{Standard Model quarks}

The SM quarks can be put together in the following table

\bigskip
\begin{tabular}{c c c c c c}
$(1/2,0)$ reps & $SU(3)$ & $SU(2)$ & $Y$ & $T_3$ & $Q=T_3+Y$ \\
\hline
$Q_i = \left(\begin{array}{c} u_i \\ d_i \end{array} \right)$ & $\begin{array}{c} {\rm triplet}  \\ {\rm triplet} \end{array}$ & doublet & $\begin{array}{c} 1/6  \\ 1/6\end{array}$ & $\begin{array}{c} 1/2  \\ -1/2\end{array}$ & $\begin{array}{c} 2/3  \\ -1/3\end{array}$ \\
$\bar{u}_i$ & triplet & singlet & $-2/3$ & 0 & $-2/3$ \\
$\bar{d}_i$ & triplet & singlet & $1/3$ & 0 & $1/3$ \\
\hline
\end{tabular}

\bigskip
All quarks here are unprimed two-component spinors. Therefore, the Hermitian conjugate of $u_i$ is denoted by $u_i^\dagger$, and $\bar{u_i}$ is another independent unprimed spinor. Notice that each $SU(n)$ multiplet corresponds to a collection of unprimed spinors. Thus, $e.g.$ $u_i$, which is a color triplet, has two types of indices suppressed: the usual spinor index, as well as the strong ${ SU}(3)$ index. With the index structure made explicit, this field is denoted by $u_{i A}^\alpha$, where $A=1,2$ is the usual spinor index, and $\alpha=1,2,3$ is the index on which ${SU}(3)$ acts. Only the flavour index $i=1,2,3$ is left explicit in what follows.

\subsection{Higgs field}

This is the field that plays the central role in the the Standard Model. It is a complex scalar field of ${ U}(1)$ hypercharge charge $Y=1/2$. It is also a weak ${ SU}(2)$ doublet, $i.e.$ it can be written as a column

\bigskip
\begin{tabular}{c c c c c c}
Higgs & $SU(3)$ & $SU(2)$ & $Y$ & $T_3$ & $Q=T_3+Y$ \\
\hline
$\phi = \left(\begin{array}{c} \phi^+ \\ \phi^0 \end{array} \right)$ & singlet & doublet & $\begin{array}{c} 1/2  \\ 1/2\end{array}$ & $\begin{array}{c} 1/2  \\ -1/2\end{array}$ & $\begin{array}{c} 1  \\ 0\end{array}$ \\
\end{tabular}

\bigskip
Note that being an ${SU}(2)$ doublet, it is really a collection of 2 complex scalar fields $\phi^+$ and $\phi^0$ (with $Q=0$). We shall denote the weak ${ SU}(2)$ index by $a, b, \ldots = 1,2$. Thus we can write the Higgs field as $\phi_a$, with $\phi_1=\phi^+$ and $\phi_2=\phi^0$.

\subsection{Quark sector of the Standard Model}

Using index-free notations, the Lagrangian for the quark sector of the Standard Model reads:
\begin{align}\begin{split}
\mathcal{L}_{quarks} = ~& -i\sqrt{2}Q^{\dagger i}{D} {Q_i} - i\sqrt{2}\bar u^{\dagger i}{D} \bar u_i- i\sqrt{2}\bar d^{\dagger i}{D} \bar d_i \\
&+ Y_u^{ij}  \phi^T \varepsilon Q_i \bar u_j - Y_d^{ij} \phi^\dagger Q_i \bar d_j - (Y_u^\dagger)^{ij} \bar u_i^\dagger Q^\dagger_j\varepsilon  \phi^*   - (Y_d^\dagger)^{ij}\bar d_i ^\dagger Q_j  ^\dagger\phi 
 \end{split}\label{Lferm}\end{align}

Here as before $D^{AA'} \equiv \theta^{\mu AA'}D_\mu$, where $D_\mu$ is the covariant derivative taht acts on each field according to its representation. The quantities $Y^{ij}$ are arbitrary complex $3\times 3$ Yukawa matrices. The quantity $\epsilon\equiv \epsilon_a{}^b$ is the matrix 
\begin{align}
\epsilon_a{}^b=\left(\begin{array}{cc} 0 & 1 \\-1 & 0 \end{array} \right).
\end{align}

which plays the role of a $SL(2,\mathbb{C})$ metric. Then the object $\phi^T \epsilon Q \equiv (\phi^T)^a \epsilon_a{}^b Q_a$ is invariant under the action of the latter via $Q\to gQ, \phi\to g\phi$ since $g^T \epsilon g=\epsilon$. In particular, $\phi^T \epsilon Q$ is ${SU}(2)$ invariant. It is then clear that all the interaction terms in (\ref{Lferm}) are ${SU}(2)$ invariant. The hypercharge invariance is easily checked using the charges tables.

\section{Second order formulation of the Standard Model}
\label{sec:SOL}
We now formalise the procedure of integrating out the primed two-component spinors from the significantly more involved quarks Lagrangian. The equations of motion for the unprimed spinors are:
\vspace{-25pt}
\begin{spacing}{2}
\begin{align}\begin{split}
\begin{array}{ccl}
Q_i^{\dagger}: & i\sqrt{2}D Q^i =& -\left( \epsilon\phi^*\right) \bar{u}^\dagger_j (Y_u^\dagger)^{ji}- \phi ~\bar{d}^\dagger_j (Y_d^\dagger)^{ji} \\  
\bar u_i^{\dagger}: & i\sqrt{2}D\bar u^i =& - (Y_u^\dagger)^{ij}Q^\dagger_j\left( \epsilon\phi^*\right) \\
\bar{d} _i^\dagger: &i\sqrt{2} D\bar d^i =&- (Y_d^\dagger)^{ij}Q^\dagger_j~\phi \\ 
\end{array}\end{split}
\end{align}
\end{spacing}

Notice that some symmetry structure is making itself explicit in the equations of motion. Thus, let us combine the components of the Higgs field into the following $2\times 2$ matrix:
\begin{align}
\rho\Phi^\dagger \defeq\left( \epsilon\phi^*, \phi\right ) \equiv  \left( \begin{array}{cc}
(\phi^0)^* & \phi^+ \\ - \phi^- & \phi^0\\
\end{array}\right).
\end{align}
Under the weak $SU(2)$ the matrix $\Phi^\dagger$ transforms as:
\begin{align}
\Phi^\dagger  ~\mapsto~ \Omega\Phi^\dagger,
\end{align}
while the field $\rho$ remains invariant. It is clear that $\rho^2 \equiv |\phi |^2$ is just the modulus squared of the Higgs field.

In order to further simplify the equations of motion, a series of field redefinitions are needed, \cite{Espin:2013wia}. This leads to:

\vspace{-25pt}
\begin{spacing}{2}
\begin{align}\begin{split}
\begin{array}{ccl}
 Q _i^\dagger: & i\sqrt{2}D Q_i =& -\rho~\Phi^\dagger \left( \bar{Q}^\dagger\Lambda\right)_i\\  
\bar{Q} _i^\dagger: &i\sqrt{2} D\bar Q_i =&- \rho~ Q^\dagger_i \Phi^\dagger\\ 
\end{array}\end{split}
\end{align}
\end{spacing}

Here, the doublet $Q_i$ transforms under the weak ${SU}(2)$, and so does the Higgs matrix $\Phi^\dagger$, while $\bar{Q}_i$ does not transform. This suggests that we define new composite ${\rm SU}(2)$-invariant quark variables $\Phi Q_i$
\begin{align}\label{new-Q}
\Phi Q_i \defeq Q_i^{inv}.
\end{align}
This corresponds to a Higgs-field dependent ${ SU}(2)$ gauge transformation of the original quark doublet. As such, it can be pulled through the (gauge dependent) covariant derivative if one transforms the $SU(2)$ gauge fields accordingly. The new vector field so defined will be an ${ SU}(2)$-invariant object. Notice that the $SU(2)$ invariance does not disappear from the theory, it is simply ``frozen'' by the new set of variables that we chose as a convenient basis to work in. This change of variables for the gauge field is equivalent to working in the unitary gauge, but it should be stated that the second order formulation exists independently of the choice of variables. Keeping in mind this change in the derivative operator we can write the field equations as:
\vspace{-25pt}
\begin{spacing}{2}
\begin{align}\begin{split}
\begin{array}{|c|cl|}
\hline  Q _i^\dagger: & i\sqrt{2}D Q_i =& -\rho~\left(\bar{Q}^\dagger \Lambda\right)_i \\  
\bar{Q} _i^\dagger: &i\sqrt{2} D\bar Q_i =&- \rho~ Q^{\dagger}_i \\ \hline
\end{array}\end{split}\label{realquarks}
\end{align}
\end{spacing}

where from now on we drop the superscript $inv$ from the $Q_i$ as it is clear that we will be working with this set of variables exclusively. We see that the equations become much simpler than in terms of the original fields. 

We now substitute the primed spinors obtained from the above field equations into the Lagrangian (\ref{Lferm}) and obtain the following second-order Lagrangian:
\begin{align}\label{L2-quarks}
\mathcal{L}_{quarks}= -\frac{2}{\rho}D\bar Q^i{D} {Q_i} - \rho \left( \Lambda \bar Q\right)^i Q_i ,
\end{align}

where it should be remembered that the covariant derivative acting on $Q_i$ in (\ref{L2-quarks}) takes into account the field redefinition (\ref{new-Q}). 

Notice that this new second-order Lagrangian, even though it contains fewer terms that its first-order counterpart, is clearly non-polynomical in the (physical) Higgs scalar field $\rho$, because of the presence of $1/\rho$ in the kinetic term. In the case of Dirac theory (\ref{eqs-dirac}) a constant rescaling of the spinors was all that we needed to obtain a canonical kinetic term. The same procedure can be applied to (\ref{L2-quarks}). However, $\rho$ is now a dynamical field and thus, absorbing it into the fermion fields (again) changes the covariant derivative operators acting on both $\bar{Q}_i, Q_i$. Denoting the new covariant derivative by ${\mathcal D}$, we finally write:
\begin{align}\label{L2-quarks*}
\boxed{\mathcal{L}_{quarks} =  -2\mathcal{D}\bar Q^i\mathcal{D} {Q_i} - \rho^2 \left(\Lambda \bar Q\right)^i Q_i }
\end{align}
where $1/\sqrt{\rho}$ was absorbed into each spinor field. The new covariant derivative $\mathcal{D}$ contains non-polynomial Higgs-quarks interactions as well as the physical $SU(2)$-frozen gauge fields when acting on the unbarred doublet. It is clear that interaction vertices with the Higgs can be of arbitrarily high valency (due to non-polynomiality in $\rho$). 

The field equations (\ref{realquarks}) for the new fermionic fields of mass dimension one read

\be\label{quarks-RC}
i\sqrt{2}{\mathcal D} Q_i = -\rho~\left(\bar{Q}^\dagger \Lambda\right)_i , \qquad i\sqrt{2} {\mathcal D}\bar Q_i =- \rho~ Q^{\dagger}_i.
\ee
and are now interpreted as the reality conditions, whose linearised versions are to be imposed on the external lines.

\section{Perturbation theory and perturbative Unitarity}
We now discuss, in the framework of Dirac theory, how pertubation theory is modified but leads to the same results as in the first-order formalism, and also how Unitarity holds even though the Lagrangian is non-Hermitian. More details can be found in \cite{paperuni}.

\subsection{Feynman rules}
The propagator becomes very simple:
\begin{align}
\langle 0| T\{\chi_A(p)\xi_B(-p)\}|0 \rangle \equiv D(p)_{AB} = \frac{-i}{p^2+m^2}\epsilon_{AB}
\end{align}
It is nothing but a scalar-type propagator, the espilon tensor in the numerator being the identity over the space of unprimed spinors. The complexity of the Dirac propagator is now shifted to the vertices. We have two interaction vertices with Feynman rules (incoming momenta):
\begin{align}
 2i e \left[k_C{}^{A'}\epsilon_{BA} + p_B{}^{A'}\epsilon_{CA} \right],\qquad -2ie^2 \epsilon^{A'B'}\epsilon_{AB}\epsilon_{CD}
\end{align}
Notice that the quartic vertex is simply the identity over both Minkowski spacetime and the space of unprimed spinors.

\vspace{-35pt}\begin{figure}[H]\begin{center}
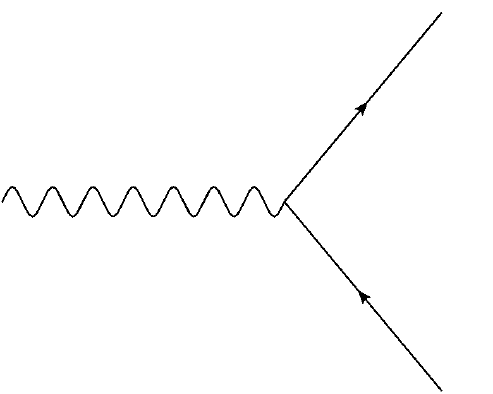 \quad 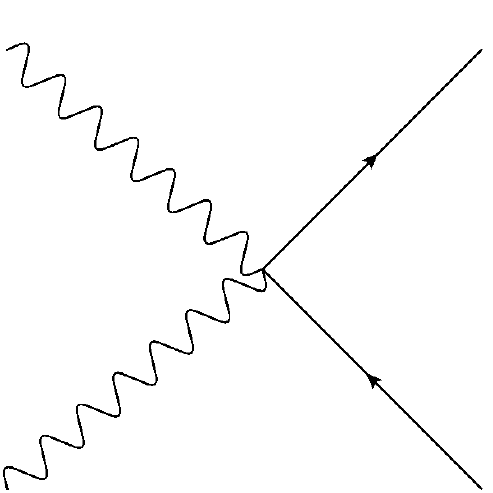
\caption{Interaction vertices}
\end{center}\end{figure}

\subsection{Spin averaged probabilities}
When we sum (or average) over photon polarisation states, one can make use of the Ward identities to obtain:

\begin{align}
\sum_{\rm pol.}\epsilon_\mu\epsilon^*_\nu ~\rightarrow ~ \eta_{\mu\nu} 
\end{align}

In our case, this will become:

\begin{align}
\sum_{\rm pol.}\epsilon_{AA'}\epsilon^*_{BB'} ~\rightarrow ~ -\epsilon_{AB}\epsilon_{A'B'}  
\end{align}

As for the second-order fermions, we have:

\begin{align}
\epsilon_A^+\epsilon_{A'}^{*+}(k) + \epsilon_A^-\epsilon_{A'}^{*-}(k) = \sqrt{\frac{2}{m^2}}k_{AA'}
\end{align}

\subsection{$e^-\mu^- \rightarrow e^-\mu^-$ scattering}

\begin{figure}[H]
	\begin{center}
	 	\includegraphics[width=0.3\linewidth]{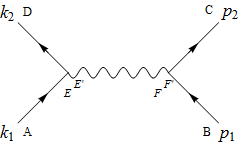}
	 	\caption{$e^-\mu^- \rightarrow e^-\mu^-$}\label{emuemu}
	\end{center}
\end{figure}

Consider the simplest QED process: electron-muon scattering at tree level in the limit $m_e \ll m_\mu$, Fig.\ref{emuemu}.

Let us first compute the amputated amplitude $\mathcal{M}_{ABCD}$ for an incoming electron with momentum $k_1$ scattered off an incoming muon with momentum $p_1$. We have:

\begin{align}\begin{split}
\mathcal{M}_{ABCD}&= 4\frac{(ie)^2(-i)}{q^2}\left(k_{1A}{}^{E'}\epsilon_{DE}-k_{2D}{}^{E'}\epsilon_{AE}\right) \epsilon^{EF}\epsilon_{E'F'}\left(p_{1B}{}^{F'}\epsilon_{CF}-p_{2C}{}^{F'}\epsilon_{BF}\right)\\ &=- \frac{4ie^2}{q^2}\Big[\left(k_1\cdot p_1\right)_{AB}\epsilon_{CD} -\left(k_2\cdot p_1\right)_{DB}\epsilon_{AC} -\left(k_1\cdot p_2\right)_{AC}\epsilon_{BD} +\left(k_2\cdot p_2\right)_{CD}\epsilon_{AB}   \Big]
\end{split}\end{align}
where we defined:

\begin{align}
(k\cdot p)_{AB} \defeq k_A{}^{C'}p_{BC'}
\end{align}
and $q^2=(k_1-k_2)^2=(p_1-p_2)^2 = t$. The complex conjugate amplitude is simply given by replacing every unprimed spinor by a primed one and every primed spinor by an unprimed one, so that:

\begin{align}
\mathcal{M}^*_{A'B'C'D'}=- \frac{4ie^2}{q^2}\Big[\left(k_1\cdot p_1\right)_{A'B'}\epsilon_{C'D'} -\left(k_2\cdot p_1\right)_{D'B'}\epsilon_{A'C'} -\left(k_1\cdot p_2\right)_{A'C'}\epsilon_{B'D'} +\left(k_2\cdot p_2\right)_{C'D'}\epsilon_{A'B'}   \Big]\label{ampprime}
\end{align}
and we defined:

\begin{align}
(k\cdot p)_{A'B'} \defeq k_{A'C}p_{B}{}^{C}
\end{align}

The unpolarised cross-section is then obtained from:

\begin{align}\begin{split}
\overline{|\mathcal{M}|^2} &= \frac{1}{4}\sum_{\rm pol.} \mathcal{M}_{ABCD}\mathcal{M}^*_{A'B'C'D'}\epsilon^A\epsilon^{*A'}(k_1)\epsilon^B\epsilon^{*B'}(p_1)\epsilon^C\epsilon^{*C'}(p_2)\epsilon^D\epsilon^{*D'}(k_2) \\ &=\frac{1}{4} \mathcal{M}_{ABCD}\mathcal{M}^*_{A'B'C'D'}k_1^{AA'}p_1^{BB'}p_2^{CC'}k_2^{DD'}\frac{2}{m_e^2}\frac{2}{m_\mu^2}
\end{split}\end{align}

With a bit of algebra, it can be shown that the following equality holds:

\begin{align}
\mathcal{M}^*_{A'B'C'D'}k_1^{AA'}p_1^{BB'}p_2^{CC'}k_2^{DD'}\frac{2}{m_e^2}\frac{2}{m_\mu^2} = -\mathcal{M}^{ABCD}
\end{align}

So that:

\begin{align}\begin{split}
\overline{|\mathcal{M}|^2}  &=- \frac{1}{4} \mathcal{M}_{ABCD}\mathcal{M}^{ABCD}
\end{split}\end{align}

In the above formula, we only need to compute three different expressions. Consider four momenta $k,~p,~q,~l$ describing massive particles. We have:

\begin{spacing}{1.5}
\begin{align}
(k\cdot p)_{AB}\epsilon_{CD} \left\lbrace \begin{array}{ll} (k\cdot p)^{AB}\epsilon^{CD} &= m_k^2 m_p^2 \\ (k\cdot q)^{AC}\epsilon^{BD} &= -\frac{1}{2}m_k^2 (p\cdot q) \\ (q\cdot l)^{CD}\epsilon^{AB} &= (k\cdot p)(q\cdot l) \end{array} \right.
\end{align}
\end{spacing}

where

\begin{align}
k\cdot p = k^{A}{}_{A'}p^{A'}{}_{A}= k_\mu p^\mu
\end{align}

Using this and neglecting terms proportional to the electron mass, we obtain:

\begin{align}
\overline{|\mathcal{M}|^2}  &=\frac{8e^4}{q^4}\Big[(k_1\cdot p_1)(k_2\cdot p_2) + (k_1\cdot p_2)(k_2 \cdot p_1) + m_\mu^2 (k_1\cdot k_2) \Big]
\end{align}

which is the well know squared amplitude for the unpolarised process.

\subsection{One-loop charge renormalisation}
Let us now look at a simple one-loop example. Although there are half the number of fermions in our theory, fermions loops are equivalent in both formalisms. Indeed, let us look at the amplitude for the charge renormalisation in the second-order formalism:

\begin{align}\begin{split}
&i\Pi^{\rm 1-loop}(k)^{A'}{}_{A}{}^{B'}{}_{B} \\ &= (-1)4e^2 \int\frac{d^4 p}{(2\pi)^4}\frac{\left[ p^{A'}{}_{B}(p+k)^{B'}{}_{A} + (p+k)^{A'}{}_{B}p^{B'}{}_{A} - \frac{1}{2}\left( (p+k)^2+p^2\right) \epsilon^{A'B'}\epsilon_{AB}  \right]}{\left[p^2+m^2 \right]\left[(p+k)^2+m^2 \right]}\\
&+ (-1)4e^2 \int\frac{d^4 p}{(2\pi)^4}\frac{\epsilon^{A'B'}\epsilon_{AB} }{\left[p^2+m^2 \right]}
\end{split}
\end{align}
 where we left explicit the contributions from both diagrams. 
 
\begin{figure}[H]\begin{center}
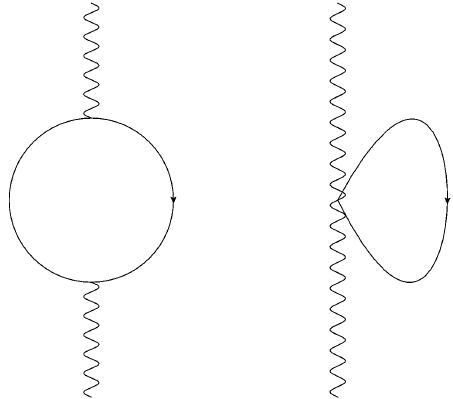
\caption{One-loop photon two-point amplitude}\label{charge}
\end{center}\end{figure}

We can rearrange the latter to obtain:
\begin{align}\begin{split}
&i\Pi^{\rm 1-loop}(k)^{A'}{}_{A}{}^{B'}{}_{B}\\ &= (-1)4e^2 \int\frac{d^D p}{(2\pi)^D}\frac{\left[ p^{A'}{}_{B}(p+k)^{B'}{}_{A} + (p+k)^{A'}{}_{B}p^{B'}{}_{A} +m^2 \epsilon^{A'B'}\epsilon_{AB}  \right]}{\left[p^2+m^2 \right]\left[(p+k)^2+m^2 \right]}
\end{split}
\end{align}

which is the usual tow-component version of the loop integral that has to be calculated in the first-order formalism.

\subsection{Remarks on perturbative Unitarity}
In quantum field theory, the S-matrix is generally written in the interaction picture:

\begin{align}
S= \mathcal{T}e^{i\int d^4x \mathcal{L}_{int}(x)}
\end{align}

where $\mathcal{T}$ denotes time-ordering. Taking matrix elements between physical states, Unitarity of the S-matrix reads:

\begin{align}
\langle f | i \rangle = \sum_{{\rm phys}~n} \langle f | S| n \rangle \langle n |S^\dagger | i \rangle = \sum_{{\rm phys}~n} \langle f | S^\dagger| n \rangle \langle n |S | i \rangle
\end{align}

where the sum runs only over physical intermediate states, which are eigenstates of the free Hamiltonian, $i.e.$ they correspond to on-shell particles. For perturbative computations, one splits the S-matrix into the identity plus a transition matrix $T$:

\begin{align}
S=1+iT
\end{align}

The Unitarity equation then reads:

\begin{align}
\langle f |(iT)| i \rangle +\langle f |(iT)^\dagger| i \rangle= -\sum_{{\rm phys}~n} \langle f | (iT)^\dagger| n \rangle \langle n |(iT)| i \rangle
\end{align}

This equation is generally proven in Perturbation Theory, order-by-order in the coupling constant appearing in the interactions, and diagram-by-diagram. Usually, this is shown to be true as a consequence of the Hermicity of the Lagrangian, however, in \cite{paperuni}, we show that this latter requirement is not necessary and we prove Unitarity of second-order QED at all orders. The proof relies on the existence of a non-trivial real-structure that involves a derivative operator: \be \dagger ~ \mapsto ~ \frac{i}{m} \mathcal{D}, \quad \left(\frac{i}{m} \mathcal{D}\right)^2 = I_V\ee

This is nothing but the first-order Dirac equation imposed on the physical states of the theory. Imposed linearly (without any coupling to the gauge fields) on the free theory, this leads to a positive definite Hamiltonian $H_0 \sim a^\dagger a$. While, imposed on the external states, they lead to a Unitary S-Matrix.

\section{Conclusion}

In this short note, we demonstrate that a second-order description of spinor fields is viable and that in many aspects, it is simpler than the usual first-order description. This new reformulation of the SM leads in particular to new insights on the available Unification patterns that could be allowed \cite{Espin:2013wia}. Furthermore, since the formalism involves only half the number of spinor fields and because the interaction vertices seen as matrices live in a much smaller space than their four-dimensional counterparts, the complexity of perturbative calculations is greatly decreased. It is also believed that for the purpose of lattice calculations, working with a scalar-type propagator could simplify the implementation of the theory.

Finally, it should be noted that the construction that is followed here intends to mimic the SM calculations, but it might be possible to take the second-order formulation as fundamental and try to implement modifications of the theory in the latter.

\end{document}